\documentclass[letterpaper]{article} % DO NOT CHANGE THIS
\usepackage{aaai24}  % DO NOT CHANGE THIS  % for submission, add [submission]
\usepackage{times}  % DO NOT CHANGE THIS
\usepackage{helvet}  % DO NOT CHANGE THIS
\usepackage{courier}  % DO NOT CHANGE THIS
\usepackage[hyphens]{url}  % DO NOT CHANGE THIS
\usepackage{graphicx} % DO NOT CHANGE THIS
\usepackage{natbib}  % DO NOT CHANGE THIS AND DO NOT ADD ANY OPTIONS TO IT
\usepackage{caption} % DO NOT CHANGE THIS AND DO NOT ADD ANY OPTIONS TO IT
\usepackage{algorithm}
\usepackage{algorithmic}

\usepackage{newfloat}
\usepackage{listings}
\DeclareCaptionStyle{ruled}{labelfont=normalfont,labelsep=colon,strut=off} % DO NOT CHANGE THIS
\floatstyle{ruled}
\newfloat{listing}{tb}{lst}{}
\floatname{listing}{Listing}
\usepackage{subfigure}
\usepackage{booktabs}
\usepackage{multirow}
\usepackage{amsmath}
\usepackage{amsfonts}

\title{
Let There Be Sound: Reconstructing High Quality Speech from Silent Videos
} 
\author{
    Ji-Hoon Kim\equalcontrib,
    Jaehun Kim\equalcontrib,
    Joon Son Chung
}
\affiliations{
    Korea Advanced Institute of Science and Technology, Daejeon, Republic of Korea\\
    
    \{jh.kim, kjaehun, joonson\}@kaist.ac.kr
}

\usepackage{bibentry}
\begin{document}

\maketitle

\begin{abstract}
The goal of this work is to reconstruct high quality speech from lip motions alone, a task also known as lip-to-speech. A key challenge of lip-to-speech systems is the one-to-many mapping caused by (1) the existence of homophenes and (2) multiple speech variations, resulting in a mispronounced and over-smoothed speech. In this paper, we propose a novel lip-to-speech system that significantly improves the generation quality by alleviating the one-to-many mapping problem from multiple perspectives. Specifically, we incorporate (1) self-supervised speech representations to disambiguate homophenes, and (2) acoustic variance information to model diverse speech styles. Additionally, to better solve the aforementioned problem, we employ a flow based post-net which captures and refines the details of the generated speech. We perform extensive experiments on two datasets, and demonstrate that our method achieves the generation quality close to that of real human utterance, outperforming existing methods in terms of speech naturalness and intelligibility by a large margin. Synthesised samples are available at our demo page: \url{https://mm.kaist.ac.kr/projects/LTBS}.
\end{abstract}

\section{Introduction}
Have you ever wondered what Charlie Chaplin’s movies would sound like if they weren’t silent? Indeed, there have been many discussions about what is said in archival silent movies~\cite{smith1987when,midgley2006new}. The ability to reconstruct speech from silent videos opens up interesting applications, such as redubbing silent movies, simulating natural utterance for those who suffer from aphonia, and understanding conversations at a distance.

As a result, the research in lip-to-speech has attracted an increasing amount of attention in recent years~\cite{kumar2019lipper,mira2022svts,kim2023lip}. This line of research has also benefited from the advances in deep learning methods, and in particular, self-supervised learning, since the training leverages natural occurrence of audio and video as a mode of supervision.

A lip-to-speech (LTS) system aims to learn a mapping from silent lip movements to the corresponding speech. 
This is a challenging one-to-many mapping caused by the two major obstacles. 
One is the existence of homophenes, words that have almost identical lip movements but distinct phonemes (e.g. `bit' and `pit').
The ambiguity of homophenes brings the one-to-many relationship between lip motions and phonemes~\cite{son2017lip,kim2022distinguishing}.
Another obstacle is the multiple variations in speech; same phonemes can be mapped to diverse speech styles based on individual characteristics such as timbre, intonation, and accents \cite{elias2021parallel,kim2022fluenttts}.

Numerous attempts have been made to improve the quality of LTS systems.
Existing works~\cite{le2015reconstructing,le2017generating} utilise hand-crafted visual features. However, deep-learning based approaches employ end-to-end methods that generate auditory features directly from silent lip motions. 
Early deep-learning based methods~\cite{ephrat2017vid2speech,kumar2019lipper} estimate linear predictive coding (LPC) features within a short video clip. 
Recently, many works~\cite{prajwal2020learning, kim2021lip, yadav2021speech, he2022flow, mira2022svts, kim2023lip} adopt mel-spectrogram as a regression target because it contains more sufficient acoustic information than LPC.
Despite the advances, the previous methods do not fully address the one-to-many mapping issue, suffering from a mispronounced and over-smoothed synthetic speech.

In this paper, we propose a novel LTS system that highly improves the synthetic quality by alleviating the intrinsic one-to-many mapping problem. 
To disambiguate homophenes, we employ self-supervised speech representations as a condition for linguistic information. 
Previous studies~\cite{baevski2020wav2vec,hsu2021hubert} have proved that self-supervised learning (SSL) speech models can acquire rich speech representations without manually labeled text.
In particular, it has been demonstrated that the representations from specific layers of the SSL model contain elaborate linguistic information independent of paralinguistic features~\cite{yang21c_interspeech,chang2022distilhubert}. 
Motivated by this, we explore the intermediate layers of SSL model and utilise the hidden representations to produce accurate content without using text labels.

Moreover, we adopt acoustic variance information such as pitch and energy in order to model diverse speech variations. 
With the help of the acoustic variations, the model can not only ease the one-to-many mapping but also learn prosody of speech which is a key factor for realistic speech synthesis~\cite{skerry2018towards,sun2020fully}. 
To further address the one-to-many mapping, we use a flow based post-net~\cite{ren2021portaspeech} which refines acoustic representations with enhanced modelling capability of capturing fine-grained details~\cite{ren2022revisiting}.
Combined with the variance information, the post-net helps to learn the complex one-to-many-mapping between phonemes and speech, thereby improving the naturalness of the synthesised speech.

We conduct extensive experiments on two public datasets with qualitative and quantitative evaluation metrics. The results show that the proposed method achieves exceptional generation quality, making a mean opinion score (MOS) gap of only $\mathbf{0.28}$ in naturalness and $\mathbf{0.16}$ in intelligibility compared to the \textit{vocoded} speech\footnote{\textit{Vocoded} speech refers to speech reconstructed from the ground truth mel-spectrogram through a vocoder, and thus it is practically considered the upper bound quality for our evaluation.}.

In summary, we directly tackle the intrinsic one-to-many mapping problem of LTS, arising from the existence of homophenes and multiple speech variations.
We adopt self-supervised speech representations as a linguistic condition to disambiguate homophenes without using text labels. 
To model speech variations, we employ acoustic variance information which helps to capture diverse speech styles. 
Experimental results prove that our method outperforms existing approaches, achieving human-like generation quality.

\section{Related Works}

\subsection{Lip-to-Speech} 
With the rapid development of deep-learning, LTS pipelines have been simplified to end-to-end systems, eliminating the need for hand-crafted visual features~\cite{le2015reconstructing,le2017generating}. Vid2Speech~\cite{ephrat2017vid2speech} and Lipper~\cite{kumar2019lipper} adopt convolutions to estimate low-dimensional LPC features. However, since the LPC features contain insufficient information for producing audible waveforms, recent works mostly utilise mel-spectrogram as a regression target. Taking mel-spectrogram,~\cite{prajwal2020learning} modifies sequence-to-sequence TTS model~\cite{shen2018natural} to construct an LTS system in a single speaker scenario. VCA-GAN~\cite{kim2021lip} and \cite{yadav2021speech} further enhance the quality by taking speech variations into account.

To leverage contextual information, GlowLTS~\cite{he2022flow} proposes a non-autoregressive network architecture based on transformer~\cite{vaswani2017attention}, and SVTS~\cite{mira2022svts} adopts conformer~\cite{gulati2020conformer}.
Most recently, MT~\cite{kim2023lip} employs content supervision by using additional text labels to produce accurate content. 
In contrast to previous studies, we focus on alleviating the one-to-many mapping problem in LTS systems by clarifying homophenes and explicitly modelling speech variations, without using text labels.

\subsection{Text-to-Speech}
Text-to-speech (TTS) systems transform text inputs to intermediate acoustic representations (e.g. mel-spectrogram), which are then converted into audible waveform by a vocoder~\cite{prenger2019waveglow,yamamoto2020parallel,kong2020hifi,kim21f_interspeech}. The TTS systems have evolved over time, shifting from conventional concatenative~\cite{hunt1996unit} and statistical parametric approaches~\cite{black2007statistical} to end-to-end systems~\cite{shen2018natural,ren2020fastspeech2,lancucki2021fastpitch,lee2021multi,popov2021grad}.
While recent TTS systems achieve human-like synthetic quality, they require a large amount of manually annotated transcriptions. LTS systems have attracted increasing attention since they can be trained without transcriptions, but the generation quality largely lags behind that of TTS. In this work, we propose a high-quality LTS method that benefits from the self-supervised nature, while generating natural speech whose quality is comparable to that of TTS.

\subsection{Self-Supervised Learning}
Over the recent years, SSL has been an emerging approach for acquiring comprehensive data representations from unlabeled data, and it has achieved notable success in natural language processing~\cite{devlin2018bert} and computer vision~\cite{bachman2019learning, lin2021completer}. In speech processing, wav2vec2.0~\cite{baevski2020wav2vec} and HuBERT~\cite{hsu2021hubert} demonstrate promising results, providing various applications such as speech recognition~\cite{baevski2021unsupervised}, voice conversion~\cite{lee2021voicemixer,choi2021neural}, and speech resynthesis~\cite{polyak2021speech}. Following works~\cite{yang21c_interspeech,chang2022distilhubert} report that the hidden representations from different layers pertain to distinct speech attributes such as voice characteristics and linguistic content. Inspired by this, we leverage self-supervised hidden representations from the specific layer of the SSL model to provide linguistic condition, mitigating homophene problems.

\begin{figure*}
    \centering
    \subfigure[Overall Architecture]{
        \includegraphics[width=0.49\columnwidth]{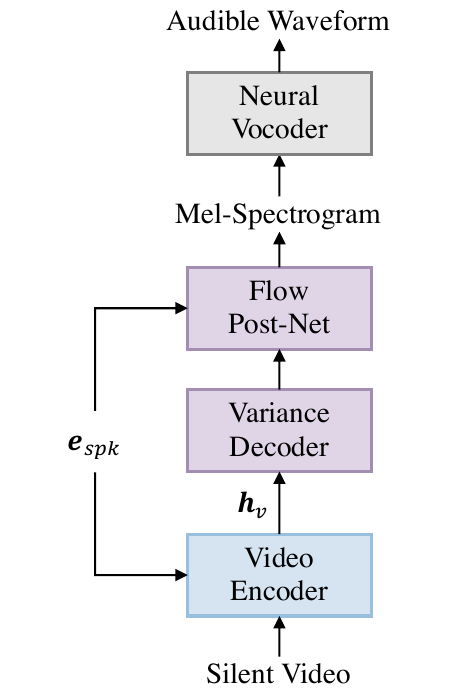}
        \label{fig:overall}}
    \subfigure[Video Encoder]{
        \includegraphics[width=0.49\columnwidth]{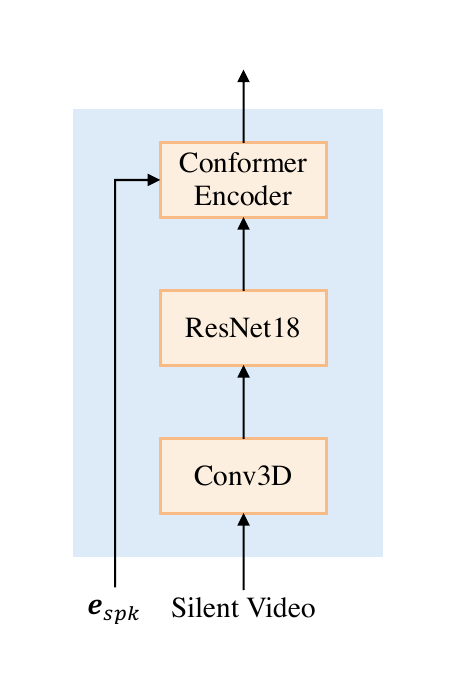}
        \label{fig:enc}}
    \subfigure[Variance Decoder]{
        \includegraphics[width=0.49\columnwidth]{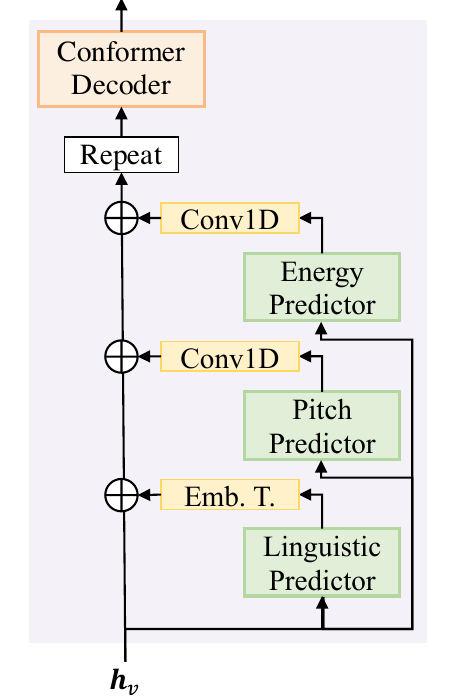}
        \label{fig:dec}}
    \subfigure[Flow Post-Net]{
        \includegraphics[width=0.49\columnwidth]{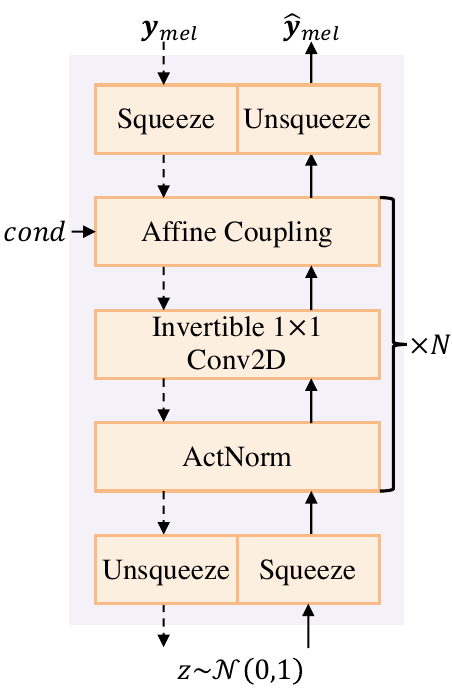}
        \label{fig:post}}
    \caption{In subfigure (a) and (b), $\boldsymbol{e}_{spk}$ is a speaker embedding. In subfigure (a) and (c), $\boldsymbol{h}_v$ denotes the encoded video feature. In (c), Emb.T. refers to an embedding table. In subfigure (d), the paths with dotted lines are operated only in a training stage. $\boldsymbol{y}_{mel}$ and $\hat{\boldsymbol{y}}_{mel}$ refer to the ground truth and predicted mel-spectrogram, respectively. $cond$ means the post-net conditions which contain the input and output of the conformer decoder, and $\boldsymbol{e}_{spk}$. In our experiment, we set $N=8$.}
    \label{fig:architecture}
    \vspace{-1mm}
\end{figure*}

\section{Method} 
Given a silent talking face video, our goal is to synthesise the corresponding mel-spectrogram. As shown in Figure~\ref{fig:overall}, the proposed model mainly consists of three components: video encoder, variance decoder, and flow based post-net. The video encoder extracts distinct visual features from input videos, and the variance decoder successively produces coarse mel-spectrogram conditioned on linguistic features and acoustic speech variations (i.e. pitch and energy). The flow based post-net elaborates the coarse mel-spectrogram into the fine-grained one, and the result is finally converted to an audible waveform by a pre-trained neural vocoder~\cite{kim21f_interspeech}.

\subsection{Video Encoder}
From the input video with $T_v$ frames, the video encoder extracts distinct hidden representations $\boldsymbol{h}_v\in\mathbb{R}^{T_v\times d}$ where $d$ denotes the hidden embedding dimension. 
As depicted in Figure~\ref{fig:enc}, the video encoder comprises 3D convolution~\cite{ji20123d}, ResNet18~\cite{he2016deep}, and a conformer encoder~\cite{gulati2020conformer}. 
Combined with 3D convolution, ResNet18 can capture adjacent contexts and local visual features, as proved in~\cite{petridis2018end,ma2022visual}. 
On the top of ResNet18, we add a conformer encoder which is a stack of conformer layers. 
Its capability to learn long-term dependencies allows to take advantage of the globality while preserving the locality obtained from the preceding convolutional layers~\cite{guo2022cmt}.
Moreover, since each individual has different visual characteristics, we inject speaker identity to the conformer encoder through an embedding table.
As validated in our experiments, the speaker embedding helps the model to extract discriminative visual features, enhancing naturalness and intelligibility of the synthesised speech.
 
\subsection{Variance Decoder}
To ease the one-to-many mapping problem in LTS, the variance decoder aims to generate acoustic representation with rich variance information. 
As shown in Figure~\ref{fig:dec}, the variance decoder consists of variance predictors and a conformer decoder.
The variance predictors are composed of linguistic, pitch, and energy predictor, each of which aims to condition the corresponding variance information into the hidden visual representations $\boldsymbol{h}_v$.
During training, we take the ground truth variance information to the hidden sequence, and use predicted value during inference.
The following conformer decoder then converts the empowered hidden visual features to intermediate acoustic representations.
In the following paragraphs, we describe the details of each predictor.

\subsubsection{Linguistic Predictor}
The presence of homophenes hinders the synthesis of intelligible speech with accurate pronunciation~\cite{ephrat2017vid2speech}.
Although the previous work~\cite{kim2023lip} attempts to address the ambiguity of homophenes by leveraging text supervision, it requires manually annotated text labels and fails to enjoy the benefits of the LTS system. 
To generate intelligible speech while preserving the self-supervised nature, we propose a linguistic predictor that disambiguates homophenes without the need for text labels.

To this end, we adopt quantised self-supervised speech representations.
We extract continuous linguistic representations from raw waveforms using pre-trained SSL speech model, namely HuBERT~\cite{hsu2021hubert}, and quantise the continuous representations for robust training\footnote{Before quantisation, the continuous features are downsampled to match the length of video by nearest-neighbor interpolation.}.
Previous studies~\cite{polyak2021speech, lakhotia2021generative, kreuk2021textless} report that the quantised speech representations from the specific layer of HuBERT contain elaborate linguistic features relevant to accurate pronunciation.
We investigate the effects of different configurations of linguistic feature extraction, and empirically find that the representations from the $12^{th}$ layer of HuBERT-LARGE\footnote{\url{https://huggingface.co/facebook/hubert-large-ls960-ft}}, quantised by $K$-means algorithm with $200$ clusters, exhibits the highest correlation with linguistic information.

Given the visual hidden features $\boldsymbol{h}_v$, the linguistic predictor aims to estimate the cluster indices of each frame.
The linguistic predictor is optimised by cross-entropy (CE) loss which can be formulated as follows:
\begin{equation}
    \mathcal{L}_{l}=\sum_{i=1}^{T_v}\text{CE}(\boldsymbol{l}_{i}, LinguisticPredictor(\boldsymbol{h}_{v,i})),
\end{equation}\\
where $\boldsymbol{l}_{i}$ and $\boldsymbol{h}_{v,i}$ are the $i^{th}$ cluster index of the target sequence and the $i^{th}$ visual representation, respectively.

\subsubsection{Pitch Predictor}
Pitch plays an important role in synthesising realistic speech with natural prosody~\cite{yasuda2019investigation,lancucki2021fastpitch}. However, the pitch exhibits multiple variations across gender, age, and emotions, exacerbating the one-to-many problem in LTS. 
To accurately capture pitch information from lip motions, we construct a pitch predictor~\cite{lancucki2021fastpitch} that estimates the pitch sequence based on the hidden visual features.

Following \cite{lancucki2021fastpitch}, we extract the ground truth pitch values from the ground truth audio through \texttt{pYIN} algorithm~\cite{mauch2014pyin}, and standardise them to have zero mean and unit variance for better sampling.
The extracted pitch values are successively downsampled to match the temporal dimension of the visual features.
The pitch predictor is optimised with L1 loss between the ground truth and predicted pitch sequence as follows:
\begin{equation}
    \mathcal{L}_{p}=\sum_{i=1}^{T_v}||\boldsymbol{p}_{i}-PitchPredictor(\boldsymbol{h}_{v,i})||_1,
\end{equation}
where $\boldsymbol{p}_{i}$ is the ground truth pitch for the $i^{th}$ video frame.

\subsubsection{Energy Predictor}
Energy represents the intensity of speech, which affects the volume and prosody of speech~\cite{bulut2007analysis}. We obtain the target energy sequence by taking the L2 norm of the mel-spectrogram along the frequency-axis~\cite{choi2021neural}.
To estimate the energy sequence from $\boldsymbol{h}_v$, we construct energy predictor~\cite{ren2020fastspeech2}, which is optimised by L1 loss between the ground truth and predicted energy sequence which can be formulated as follows:  
\begin{equation}
    \mathcal{L}_{e}=\sum_{i=1}^{T_v}||\boldsymbol{e}_{i}-EnergyPredictor(\boldsymbol{h}_{v,i})||_1,
\end{equation}
where $\boldsymbol{e}_{i}$ refers to the $i^{th}$ target energy value.

Each variance information is encoded into variance embeddings either through an embedding table (linguistic), or a single 1D convolution layer (pitch and energy). 
The variance embeddings are added to the visual representations $\boldsymbol{h}_v$, and the adapted representation is upsampled to match the time resolution of target mel-spectrogram.
Lastly, the conformer decoder converts the adapted representations to a coarse mel-sepctrogram.
We apply L1 loss between the ground truth mel-spectrogram and the predicted mel-spectrogram:
\begin{equation}
    \mathcal{L}_{mel}=\sum_{j=1}^{T_m}||\boldsymbol{Y}_{j}-\boldsymbol{\hat{Y}}_{j}||_1,
\end{equation}
where $Y_{j}$ denotes the $j^{th}$ frame of the ground truth mel-spectrogram with length ${T_m}$, and $\hat{Y}_{j}$ stands for the $j^{th}$ frame of predicted mel-spectrogram. 

Note that the variance predictors simplify the acoustic target distribution by providing conditional information, thereby mitigating the one-to-many mapping issue~\cite{ren2022revisiting}.
We analyse the effect of variance information in our experiment section.

\subsection{Post-Net}
Natural human speech comes with dynamic variations. However, simple reconstruction loss (L1 or L2 loss) is limited to capture such details, resulting in a blurry and over-smoothed synthetic speech~\cite{liu2022diffsinger}. 
To further improve the sample quality, we apply a flow based post-net~\cite{ren2021portaspeech} which elaborates the coarse-grained mel-spectrogram into a fine-grained one.

The architecture of the post-net is depicted in Figure~\ref{fig:post}. In training stage, the post-net transforms mel-spectrogram training data $\boldsymbol{x}$ into a tractable prior distribution through a series of invertible functions $\mathbf{f}=\mathbf{f}_0\circ\mathbf{f}_1\circ\cdots\mathbf{f}_k$, conditioned with the input and output of the conformer decoder, and the speaker embedding. The post-net is optimised with minimising the negative likelihood of data $\boldsymbol{x}$ as follows:
\begin{equation}
    \log p_{\theta}(\boldsymbol{x}) = \log p_{\theta}(\boldsymbol{z}) + \sum_{i=1}^{k}\log |\det(\boldsymbol{J}(\mathbf{f}_i(\boldsymbol{x})))|,
\end{equation}
\begin{equation}
    \mathcal{L}_{post} = - \log p_{\theta}(\boldsymbol{x}),
\end{equation}
where $p_{\theta}(\boldsymbol{z})$ is the tractable prior (isotropic multivariate Gaussian) over latent variable $\boldsymbol{z}$, and $\boldsymbol{J}$ is the Jacobian. During inference stage, we take samples $\boldsymbol{z}$ from the prior distribution and feed them into the post-net reversely to generate the final mel-spectrogram. As proved in~\cite{ren2022revisiting}, this flow-based module enhances the capability of modelling complex data distributions, which helps to address one-to-many mapping problem.

To summarise, the final loss ($\mathcal{L}_{final}$) is given by:
\begin{equation}
    \mathcal{L}_{final} = \mathcal{L}_{mel} + \lambda_{var}\mathcal{L}_{var} + \lambda_{post}\mathcal{L}_{post},
\end{equation}
where $\mathcal{L}_{var}=\mathcal{L}_{l}+\mathcal{L}_{p}+\mathcal{L}_{e}$. In our experiments, we set $\lambda_{var} = \lambda_{post} = 0.1$.

\subsection{Neural Vocoder}
To convert the predicted mel-spectrogram into an audible waveform, we utilise pre-trained Fre-GAN~\cite{kim21f_interspeech} as our neural vocoder. Based on adversarial networks, it produces frequency-consistent waveform by adopting discrete wavelet transform~\cite{daubechies1988orthonormal} as a lossless downsampling method. Compared to Griffin-Lim algorithm~\cite{griffin1984signal} which is mostly adopted in previous works~\cite{kim2021lip,he2022flow,kim2023lip}, the Fre-GAN vocoder shows exceptional performance in reconstructing waveforms.
The vocoder is used only at inference process.

\section{Experimental Settting}
\subsection{Datasets}
\subsubsection{GRID}~\cite{cooke2006audio} is one of the established datasets for lip to speech synthesis in a constrained environment. 
It contains 33 speakers and 50 minutes of short video clips per speaker.
The number of vocabularies in GRID is only 51 and the speakers always face forward with nearly no head movement. 
In our experiment, the dataset is split into train (80\%), validation (10\%), and test set (10\%) by sampling equally from all speakers.

\subsubsection{Lip2Wav}~\cite{prajwal2020learning} is a large-scale benchmark dataset for an unconstrained and large vocabulary lip to speech synthesis. It comprises real-world lecture recordings featuring 5 distinct speakers, with about 20 hours of video for each speaker. Our experiment is conducted on 2 speakers, Lip2Wav-Chemistry and Lip2Wav-Chess, as in GlowLTS~\cite{he2022flow}. The two speakers are trained jointly in a multi-speaker setting, and both are equally divided into 80-10-10\% for train, validation, and test sets.

\subsection{Preprocessing}
Audio data are resampled to 16kHz and transformed to mel-spectrograms with 40ms window length, 10ms hop length, and 80 mel filterbanks. 
To achieve the temporal synchronisation with the audio, the video data are resampled to 25 frames per second, resulting in a fixed ratio of 1 to 4 between the lengths of video and audio.
We then extract 68 face landmarks for each video frame using FaceAlignment\footnote{\url{https://github.com/ladrianb/face-alignment}}. 
Based on the landmarks, the lip regions are aligned to a fixed position, and cropped to their centers with a dimension of $112 \times 112$.
The cropped images are converted to grayscale.

\begin{table*}[t]
\centering
% \resizebox{\textwidth}{!}
{
\begin{tabular}{lcccccccc}
\toprule
\multirow{2}{*}{Method}     & \multicolumn{4}{c}{\bfseries GRID} & \multicolumn{4}{c}{\bfseries Lip2Wav} \\ \cmidrule(lr){2-5}\cmidrule(lr){6-9}
      & Nat. $\uparrow$   & Intel. $\uparrow$   &WER $\downarrow$ &CER $\downarrow$ & Nat. $\uparrow$  &Intel. $\uparrow$  &WER $\downarrow$  &CER $\downarrow$   \\ \cmidrule(lr){1-9}
Ground Truth    &$4.82\pm0.04$ &$4.82\pm0.05$  &$12.20$ &$~7.03$ &$4.80\pm0.05$   & $4.78\pm0.05$   &~$4.12$ &~$2.58$      \\
Vocoded &$4.74\pm0.05$       &$4.79\pm0.05$    &$12.44$ &~$7.18$ &$4.63\pm0.06$     &$4.73\pm0.06$ &~$6.05$ &~$4.19$    \\\cmidrule(lr){1-9}
VCA-GAN &$3.46\pm0.07$ &$4.10\pm0.08$ &$17.62$ &~$9.55$ &$2.05\pm0.08$ &$2.71\pm0.10$ &$48.73$ &$32.51$   \\
SVTS &$3.35\pm0.09$       &$3.97\pm0.09$ &$23.30$ &$13.12$ &$1.77\pm0.07$ & $2.18\pm0.10$   &$61.09$ &$41.01$ \\
MT &$2.42\pm0.09$  & $3.08\pm0.12$  &$30.56$ &$18.14$ &N/A &N/A   &N/A &N/A \\ 
\textbf{Ours}  &$\mathbf{4.46}\pm\mathbf{0.07}$ &$\mathbf{4.63}\pm\mathbf{0.07}$ &$\mathbf{17.07}$ &~$\mathbf{9.17}$ & $\mathbf{4.15}\pm\mathbf{0.08}$ &$\mathbf{3.69}\pm\mathbf{0.10}$  &$\mathbf{34.71}$ &$\mathbf{22.57}$ \\
\bottomrule
\end{tabular}
}
\vspace{-1mm}
\caption{Evaluation results. MOS results are presented with $95\%$ confidence interval. `Nat.' and `Intel.' represent MOS for naturalness and intelligibility, respectively. Note that MT~\cite{kim2023lip} cannot be trained on the Lip2Wav dataset since the model requires text transcription to be trained. $\uparrow$ denotes higher is better, $\downarrow$ denotes lower is better.}
\label{table:compare}
\end{table*}

\begin{figure*}[!t]
    \centering
    \subfigure[Ground Truth]{
    \includegraphics[height=2.351cm]{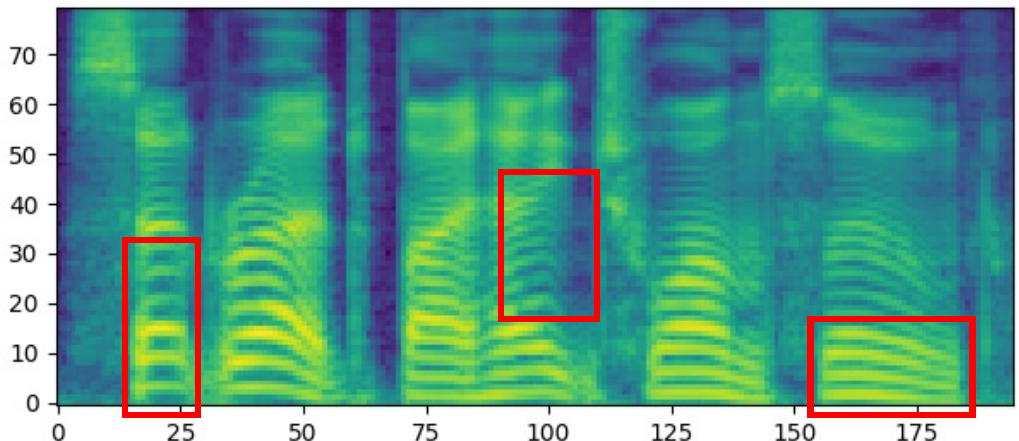}
    \label{fig:dif-frev1}}
    \subfigure[Vocoded]{
    \includegraphics[height=2.351cm]{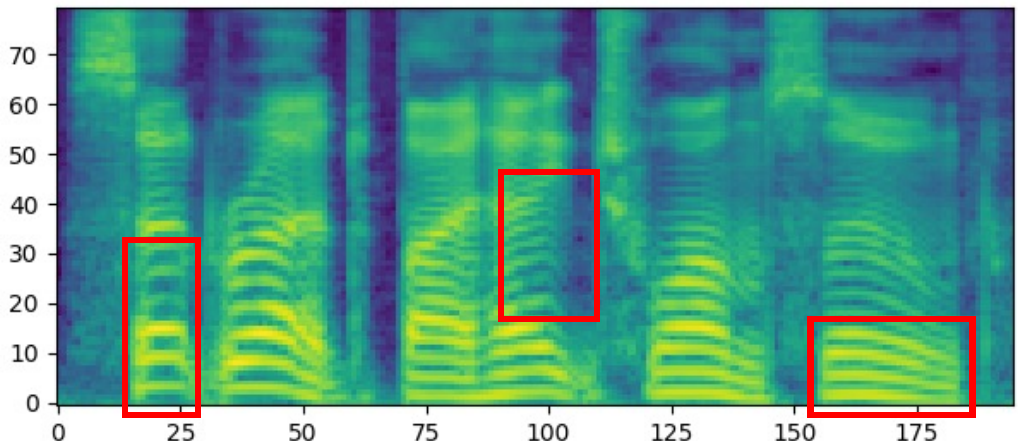}
    \label{fig:dif-hiv1}}
    \subfigure[VCA-GAN]{
    \includegraphics[height=2.351cm]{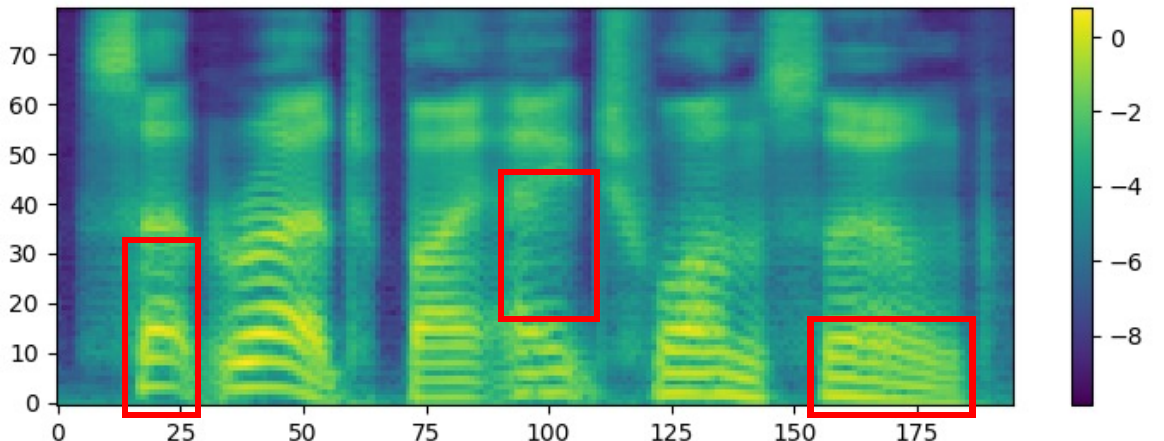}
    \label{fig:dif-wavenet}}
    
    \subfigure[SVTS]{
    \includegraphics[height=2.351cm]{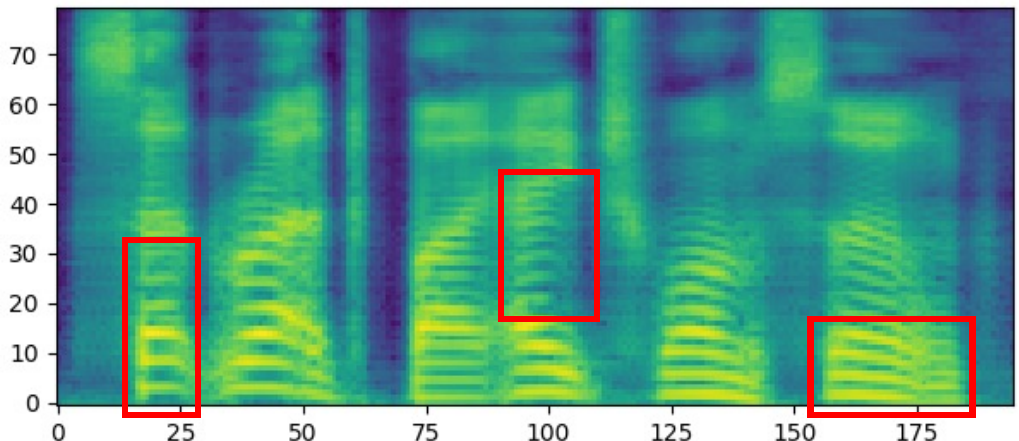}
    \label{fig:dif-frev2}}
    \subfigure[MT]{
    \includegraphics[height=2.351cm]{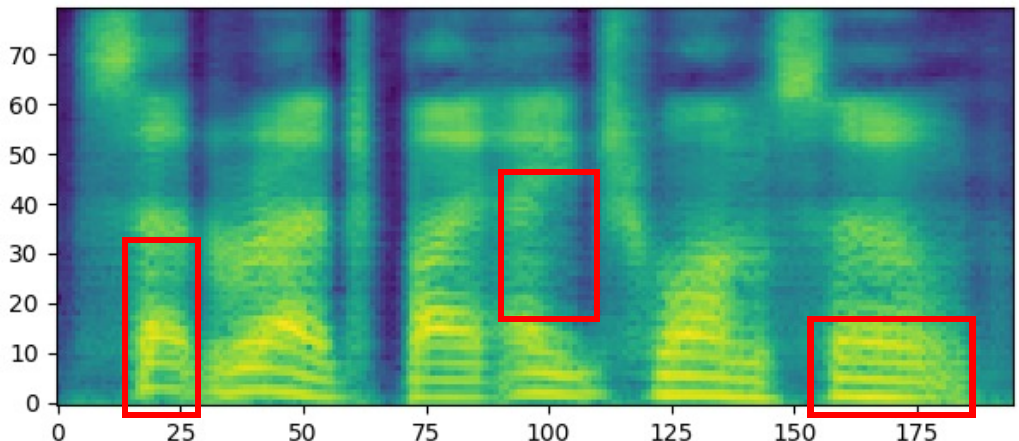}
    \label{fig:dif-hiv2}}
    \subfigure[\textbf{Ours}]{
    \includegraphics[height=2.351cm]{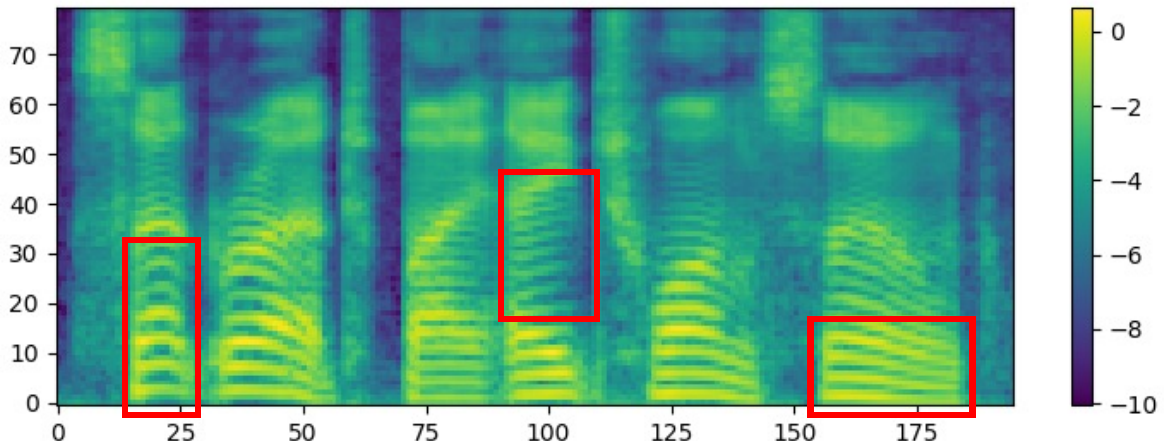}
    \label{fig:dif-waveglow}}
    \caption{Visualisation of mel-spectrogram. Note that the proposed method better captures fine details with frequency correlations compared to other methods ((c)-(e)), particularly in red boxes.}
    \vspace{-1mm}
\label{fig:mel-dif}
\end{figure*}

\subsection{Model Configuration}
We construct the identical model architecture for each dataset, with the exception of the conformer encoder.
For the constrained GRID dataset, the conformer encoder is designed with 6 attention heads and a hidden dimension of 384, and for the unconstrained Lip2Wav, the encoder is designed with 8 heads and a hidden dimension of 512.
We follow the recent works~\cite{mira2022svts, kim2023lip} for the configuration of 3D convolution and ResNet18. The speaker embedding is obtained from an embedding table, where each speaker identity is converted into a fixed size embedding vector. The pitch and energy predictors are composed of two 1D convolution layers~\cite{lancucki2021fastpitch}, while the linguistic predictor consisted of four 1D convolutions.

\subsection{Training Details}

Our model is trained on four NVIDIA A5000 GPUs with a batch size of $64$. We use AdamW optimiser~\cite{loshchilov2017decoupled} with $\beta_1=0.9$, $\beta_2=0.98$, and $\epsilon=10^{-9}$. The learning rate is fixed to $2\times10^{-4}$, and the weight decay is set to $10^{-6}$. 
For training the GRID dataset, we randomly sample consecutive sequence with a length of $50$, and the model is trained for $400$ epochs. For the Lip2Wav dataset, we sample contiguous $75$ frames and train our model for $900$ epochs.
To prevent overfitting, we apply data augmentations: horizontal flipping with probability of $50\%$, and random masking with fixed position throughout all frames. The masked area is randomly sampled within the range from $10\times10$ to $30\times30$.

\subsection{Evaluation Metrics}

The performance of the proposed method is evaluated with both qualitative and quantitative evaluation metrics. 
For qualitative evaluation, we conduct mean opinion score (MOS) test, wherein $30$ domain-expert speakers assess the quality of $30$ random speech clips for naturalness and intelligibility on a scale of $1$ to $5$.
Naturalness of speech represents how close the speech is to that of human utterance. 
Intelligibility focuses solely on the successful delivery of linguistic contents; high scores are given if one can clearly identify the contents even if it sounds unnatural. 
Moreover, we also compute word error rate (WER) and character error rate (CER) of $300$ random test samples for quantitative evaluation.
For error calculation, we obtain the transcriptions of speech clips by using publicly available automatic speech recognition (ASR) model~\cite{radford2023robust} pre-trained on 438k hours of English corpus.

Based on the above evaluation metrics, we compare our model against the ground truth, vocoded, and generated samples from recent LTS models which show promising results: VCA-GAN~\cite{kim2021lip}, SVTS~\cite{mira2022svts}, and MT~\cite{kim2023lip}. 
We follow the official implementation for VCA-GAN\footnote{\url{https://github.com/ms-dot-k/Visual-Context-Attentional-GAN}} and MT\footnote{\url{https://github.com/ms-dot-k/Lip-to-Speech-Synthesis-in-the-Wild}}. For SVTS, since there is no official implementation, we reproduce it based on the official code of MT which takes SVTS as its backbone network.
For a fair comparison, all the LTS models are trained on the same settings, and the predicted mel-spectrograms are converted to audible speech by Fre-GAN vocoder~\cite{kim21f_interspeech}.

\section{Experimental Results}
We evaluate our method in qualitative and quantitative manner, and investigate each variance prediction pipeline. Further, the effectiveness of each component in our model is verified by the ablation study. The evaluations underscore a notable enhancement over the other LTS models, and each module provides distinct contribution to increasing the quality of synthesised speech.

\subsection{Qualitative Results}
To evaluate the quality of the generated speech, we conduct MOS for naturalness and intelligibility, and the results are presented in Table~\ref{table:compare}.
For the GRID dataset, our proposed method achieves the highest naturalness and intelligibility among all generated samples.
Especially, the generated speech of the proposed method closely approximates the vocoded quality with a minor gap of $0.28$ in naturalness and $0.13$ in intelligibility.
For Lip2Wav, the overall generation quality degrades due to the unconstrained nature of the dataset.
However, the proposed method produces speech with promising quality, outperforming the existing methods by a significant margin.
Note that MT~\cite{kim2023lip} is not applicable to Lip2Wav experiment, since it requires text information for training while the dataset does not provide text labels.

Moreover, to intuitively demonstrate the effectiveness of the proposed method, we conduct mel-spectrogram visualisation analysis.
In Figure~\ref{fig:mel-dif}, we depict the generated mel-spectrograms along with the ground truth and vocoded mel-spectrogram.
Especially in red boxes, our model produces a detailed and sharp mel-spectrogram with distinct harmonics, showing close resemblance to the ground truth mel-spectrogram.
However, other methods ((c)-(e)) suffer from blurry and over-smoothed results. This indicates that our method can effectively learn the complex one-to-many mapping function, which consequently lead to natural and intelligible synthetic speech.

\subsection{Quantitative Results}
As a quantitative evaluation, we compare the WER and CER of the synthesised speech with those of the ground truth and vocoded speech.
For the GRID dataset, the error rates are obtained by directly comparing the ASR transcriptions with the provided ground truth texts.
For the Lip2Wav dataset, since the dataset does not provide text labels, we manually annotate the ground truth texts and compare them with the ASR transcription results. 

The results are shown in Table~\ref{table:compare}. 
The proposed model clearly shows the lowest WER and CER on both GRID and Lip2Wav datasets. 
This demonstrates our method can synthesise highly intelligible speech by effectively reducing the homophene problems.
Despite the higher error rates in the Lip2Wav dataset compared to GRID dataset, the proposed method achieves significantly lower WER and CER compared to all the other models, making a gap larger than 10\% point.
This explicitly supports that the proposed model is readily applicable to the unconstrained environments.

\subsection{Analysis on Acoustic Variance Information} 

To verify the effectiveness of the acoustic variance conditions, we examine the similarities between $300$ pairs of the synthesised and ground truth speech.
For pitch, we compute the moments of pitch distribution (mean ($\mu$), standard deviation ($\sigma$), skewness ($\gamma$), and kurtosis ($\kappa$)) and analyse on how much the values resemble those of the ground truth.
The results are shown in Table \ref{table:pitch}.
Each of the four values from the output of the proposed model stands the closest to those from the ground truth speech, especially the skewness value deviating only by $0.042$. This demonstrates that our model can generate speech with highly accurate pitch contour.
With the absence of the pitch predictor, the kurtosis shows a slight deviance. 
However, closing the gap of standard deviation with minimum changes in mean, skewness and kurtosis clearly supports that the pitch predictor explicitly contributes to producing high-quality results.

For energy, we calculate the frame-wise mean absolute error (MAE) between energy of generated speech and that of ground truth speech.
As shown in Table~\ref{table:mae}, the MAE from the proposed model reports the lowest among all other models by a distinct margin. 
This implies that the energy of the generated speech from the proposed model closely resembles that from the ground truth speech.
The influence of the energy predictor is also confirmed by the increase of MAE when the predictor is removed.

\begin{table}
\centering
\resizebox{0.99\columnwidth}{!}{
\begin{tabular}{lcccc}
\toprule
Method
      & $\mu$ &$\sigma$   &$\gamma$  &$\kappa$    \\  \cmidrule(lr){1-5}

Ground Truth  &$77.90$ &$101.84$ &$0.696$  &$-1.217$     \\ \cmidrule(lr){1-5}
    VCA-GAN  &$102.14$ &$95.98$ & $0.124$ & $-1.454$ \\
    SVTS  &$98.85$ &$97.95$ &$0.198$ &$-1.518$ \\ 
    MT  &$83.42$ &$91.82$ &$0.450$ &$-1.347$ \\ \cmidrule(lr){1-5}
    \textbf{Ours}  &$\mathbf{79.74}$ &$\mathbf{101.63}$ &$\mathbf{0.654}$ &$-1.284$ \\  
    ~~\textit{w/o} pitch  &$79.81$ &$102.39$ &$0.652$ &$\mathbf{-1.282}$  \\
\bottomrule
\end{tabular}
}
\vspace{-1mm}
\caption{Mean ($\mu$), standard deviation ($\sigma$), skewness($\gamma$), and kurtosis($\kappa$) of the pitch distribution for ground truth and synthesised audio.}
\label{table:pitch}
\end{table}

\begin{table}
\centering
\resizebox{0.99\columnwidth}{!}{
\begin{tabular}{l|ccc|cc}
\toprule
\multirow{2}{*}{Method} & \multirow{2}{*}{VCA-GAN} &\multirow{2}{*}{SVTS} &\multirow{2}{*}{MT} &\multirow{2}{*}{\textbf{Ours}} & \multirow{2}{*}{\begin{tabular}[c]{@{}l@{}}\textbf{Ours}\\ ~~\textit{w/o} E.\end{tabular}} \\
                         & & & & &                                                                \\ \hline
 \rule{0pt}{2.57ex} MAE $\downarrow$   & $4.155$                & $4.275$        &$5.314$ &$\mathbf{3.886}$ &$3.959$                                                    \\ \bottomrule
\end{tabular}
}
\vspace{-1mm}
\caption{The MAE between the energy of ground truth and that of synthesised audio. ``E." stands for energy.}
\vspace{-2mm}
\label{table:mae}
\end{table}

\subsection{Analysis on Self-Supervised Features}
The significance of SSL speech models is proven by the recent studies~\cite{baevski2020wav2vec,hsu2021hubert}, and the research is further explored with the utilisation of intermediate representations on various downstream tasks~\cite{yang21c_interspeech,chang2022distilhubert}. 
Outputs from the first layer are used to extract speaker identity in \cite{fan2020exploring,choi2021neural}, and \cite{lakhotia2021generative,lee2022hierspeech} utilise the middle layer to obtain linguistic representation.
Particularly, \cite{lakhotia2021generative} reports that the number of $K$-means clusters clearly affects the model performance when using the quantised linguistic representation.

To find the optimal linguistic feature configuration for our model, we compute WER, CER, and phoneme error rate (PER) on the GRID validation set using various feature extraction settings. 
To be specific, linguistic features are obtained from $1^{st}$, $12^{th}$, and $24^{th}$ layer outputs of HuBERT. The continuous linguistic features are then quantised with $100$, $200$ clusters, and the cluster indices are used as targets for the linguistic predictor.
Table~\ref{table:SSL} demonstrates that the outputs from $12^{th}$ layer of HuBERT quantised with $200$ clusters produce the most intelligible speech, achieving the lowest PER and CER. 
While the same configuration but $100$ clusters achieves the lowest WER, larger number of clusters shows lower PER and CER. Considering the fact that phoneme accuracy is closely related to the accurate pronunciation~\cite{zhang2022mixed}, the configuration with the lowest PER generates the most intelligible speech. 

\subsection{Ablation Study}
\label{sec:ablation}

To verify the effect of each module in the proposed method, we conduct an ablation study on the Lip2Wav dataset using $7$-scale comparative MOS (CMOS), WER, and CER.
In CMOS, 30 domain experts listen to the audio samples from two systems and compare the quality from -$3$ to +$3$.

As shown in Table~\ref{table:ablation}, the results of the ablation study clearly support that each component independently contributes to improving the quality of the synthetic speech.
Notably, the absence of the linguistic predictor results in the largest quality degradation in speech intelligibility, WER, and CER. This proves the effectiveness of the linguistic predictor in clarifying homophenes, which connects to speech generation with accurate pronunciation.
The significance of the acoustic variance information, especially pitch, is validated by the quality degradation when such information is not considered.
Removing the post-net shows the largest decrease in naturalness, highlighting the effectiveness of the module in producing fine details of acoustic features.
The importance of speaker information $\boldsymbol{e}_{spk}$ is proven by the degraded quality when the information is excluded.

\begin{table}
\centering
% \resizebox{0.99\columnwidth}{!}
{
\begin{tabular}{ccccc}
\toprule
\#clusters 
      &layer   &WER $\downarrow$  &PER $\downarrow$ &CER $\downarrow$    \\  \cmidrule(lr){1-5}

    100  &1       &$18.02$ &$~~9.58$  &$10.04$     \\ 
    100  &12 &$\mathbf{16.53}$ & $10.77$ & $11.39$ \\
    100  &24 &$17.57$ &$~~8.92$ &$10.10$ \\ \cmidrule(lr){1-5}
    200  &1  &$17.62$ &$~~9.73$ &$~~9.76$ \\
    200  &12 &$17.12$ &$~~\mathbf{8.91}$ &$~~\mathbf{9.70}$ \\  
    200  &24 &$29.17$ &$~~9.59$ &$16.03$  \\
\bottomrule
\end{tabular}
}
\vspace{-1mm}
\caption{Evaluation on different configurations of linguistic feature extraction. \#clusters denotes the number of $K$-means clusters and layer means the layer index of HuBERT.}
\label{table:SSL}
\end{table}

\begin{table}
\centering
\resizebox{0.99\columnwidth}{!}{
\begin{tabular}{lcccc}
\toprule
Method 
      & Nat. $\uparrow$   & Intel. $\uparrow$ &WER $\downarrow$ &CER $\downarrow$    \\ \cmidrule(lr){1-5}
\textbf{Ours} &~~~$0.00$ &~~~$0.00$  &$34.71$ &$22.57$    \\ 
    \cmidrule(lr){1-5}
    \emph{~w/o} linguistic  &$-0.90$  &$-0.70$  &$42.51$ &$27.99$\\
    \emph{~w/o} pitch  &$-1.06$ &$-0.61$ &$39.96$ &$26.30$\\
    \emph{~w/o} energy &$-0.42$ &$-0.62$  &$40.58$ &$26.46$\\
    \emph{~w/o} post-net      &$-1.48$ &$-0.57$  &$40.05$ &$25.48$\\  
    \emph{~w/o} $\boldsymbol{e}_{spk}$      &$-0.48$ &$-0.56$   &$42.33$ &$27.04$\\
\bottomrule
\end{tabular}
}
\vspace{-1mm}
\caption{CMOS, WER, and CER results of an ablation study.}
\vspace{-2mm}
\label{table:ablation}
\end{table}

\section{Conclusion}

In this paper, we propose a novel LTS system that generates speech close to human-level quality in both naturalness and intelligibility. 
We directly tackle the inherent one-to-many mapping problems of LTS, and address them by providing linguistic and acoustic variance information.
We further refine the generated speech by enhancing modelling capability.
Both qualitative and quantitative experiments clearly demonstrate that the proposed method improves the overall quality of the synthesised speech, outperforming the previous works by a notable margin.
We also verify the effectiveness of each proposed component through the ablation study, and analyse the effect of the variance information from various perspectives.
For the future work, we will continue to enhance the generated speech quality by adopting audio-visual SSL models.
We also aim to simplify the overall generation pipeline with the inclusion of neural vocoder, making a fully end-to-end architecture. 

\clearpage

\section{Broader Impact}
By employing the proposed LTS system, numerous positive societal impacts can be realised, including the dubbing of silent videos and the simulation of natural utterances for individuals with speech impairments.
However, alongside these advantages, there exist potential threats associated with the misuse of our system, such as the generation of fake speech and voice phishing.
Furthermore, as the LTS system enables one to comprehend conversations from a distance, there is a risk of its use in invading personal privacy.

\section{Acknowledgements}
This work was supported by the National Research Foundation of Korea (NRF) grant funded by the Korea government (MSIT) (No. RS-2023-00212845, Multi-modal Speech Processing for Human-Computer Interaction).

\bibliography{aaai24}

\end{document}